\documentclass[mathleft
]{an}
\usepackage{graphicx}
\usepackage{psfig}

\usepackage{times}
\usepackage{natbib}
\newcommand\aap{A\&A}
\newcommand\nat{Nature}

\newcommand\aapr{A\&A Rev.}

\begin{document}

\Pagespan{789}{}
\Yearpublication{2014}%
\Yearsubmission{2013}%
\Month{11}%
\Volume{999}%
\Issue{88}%

\title{The loss-limited electron energy in SN 1006: effects of the shock velocity and of the diffusion process}

\author{M. Miceli\inst{1}\fnmsep\thanks{Corresponding author:
  \email{miceli@astropa.unipa.it}\newline}
\and 
          F. Bocchino\inst{1}
	  \and
	  A. Decourchelle\inst{2}
	  \and
	  J. Vink\inst{3}
	  \and
	  S. Broersen\inst{3}
          \and
	  S. Orlando\inst{1}
	  }

\titlerunning{The loss-limited electron energy in SN 1006}
\authorrunning{M. Miceli et al.}
\institute{
INAF-Osservatorio Astronomico di Palermo, Piazza del Parlamento 1, 90134 Palermo, Italy
\and 
Service d'Astrophysique/IRFU/DSM, CEA Saclay, Gif-sur-Yvette, France
\and 
Astronomical Institute ``Anton Pannekoek", University of Amsterdam, P.O. Box 94249, 1090 GE Amsterdam, The Netherlands}

\received{}
\accepted{}
\publonline{later}


\abstract{%
The spectral shape of the synchrotron X-ray emission from SN 1006 reveals the fundamental role played by radiative losses in shaping the high-energy tail of the electron spectrum. We analyze data from the \emph{XMM-Newton} SN 1006 Large Program and confirm that in both nonthermal limbs the loss-limited model correctly describes the observed spectra. We study the physical origin of the observed variations of the synchrotron cutoff energy across the shell. We investigate the role played by the shock velocity and by the electron gyrofactor. We found that the cutoff energy of the syncrotron X-ray emission reaches its maximum value in regions where the shock has experienced its highest average speed. This result is consistent with the loss-limited framework. We also find that the electron acceleration in both nonthermal limbs of SN 1006 proceeds close to the Bohm diffusion limit, the gyrofactor being in the range $\eta\sim1.5-4$. We finally investigate possible explanations for the low values of cutoff energy 
measured in thermal limbs.}
\maketitle

\section{Introduction}

Synchrotron emission is a typical feature of supernova remnants (SNRs). The characteristic radio-bright shell observed in the majority of galactic SNRs (listed by \citealt{gre09}) is commonly associated with ultrarelativistic electrons accelerated up to GeV energies at the expanding shock front. The discovery of nonthermal X-ray emission in SN 1006 by \citet{kpg95} has clearly shown that diffusive shock acceleration can boost the electrons up X-ray emitting energies (as first predicted by \citealt{rc81}). This scenario has further been confirmed by the detection of X-ray synchrotron emission in other young galactic SNRs \citep{rey08,vin12}.

The presence of high-energy electrons in the shock front of SNRs suggests that also hadron acceleration can be at work. There is an emerging evidence of ultrarelativistic ions in young SNRs, e.~g., Tycho (\citealt{ekh11,mc12}) and RCW 86 (\citealt{hvb09}). 
Effects of hadron acceleration have been detected in the southeastern limb of SN 1006 by \citet{mbd12}, who found an increase in the shock compression ratio in the regions of the thermal limb that are closer to the northeastern and southwestern nonthermal limbs. This indicates the presence of shock modification induced by hadron acceleration at the shock front. More recently, suprathermal hadrons have been revealed also in the northwestern limb of SN 1006, by studying the variations in the $H\alpha$ broad line widths and the broad-to-narrow line intensity ratios \citep{nvh13}.

At odds with hadrons, X-ray emitting electrons are expected to suffer significant radiative losses via synchrotron emission. The timescale of synchrotron cooling is $t_{sync}=12.5E^{-1}_{100}B^{-2}_{100}$ yr, where $E_{100}$ is the electron energy in units of 100 TeV and $B_{100}$ is the magnetic field in units of 100 $\mu$G \citep{lon94}. The value of the magnetic field in young SNRs can be estimated from the narrowness of the X-ray synchrotron filaments (that are much broader in the radio band) and is of the order of $B_{100}\sim1-6$ (e. g., \citealt{bal06}, \citealt{vl03}, \citealt{pmb06}). These values of $B$ imply that $t_{sync} < 100$ yr for X-ray emitting electrons with $E_{100}\sim 0.1$. Also, the thinness of the X-ray shell in young SNRs clearly shows that high-energy electrons cool down very rapidly behind the shock front. Finally, in young SNRs the synchrotron cooling time is expected to be comparable to the acceleration time scale for electrons emitting X-ray synchrotron radiation (see \citealt{
mbd13}, hereafter M13, for the case of SN 1006). It is then possible that the maximum energy achieved by electron in the acceleration process is limited by their radiative losses (i. e., it is loss-limited, see \citealt{rey08}). Two alternative mechanisms can also be invoked to limit the maximum electron energy, such as limited acceleration time available, and abrupt increase in the diffusion coefficient due to the change in the availability of MHD waves above some wavelength (i. e., the time limited, and escape limited scenarios, respectively, see \citealt{rey08}).

The loss-limited scenario has been successfully adopted to describe the global (i. e., extracted from the whole SNR) X-ray spectrum of RX J1713.7-3946 (\citealt{za10}, \citealt{tua08}, \citealt{uat07}) and of Tycho \citep{mc12}.
Recently, M13 analyzed the deep observations of the \emph{XMM-Newton} SN 1006 Large Program and performed spatially resolved spectral analysis of a set of small regions in the southwestern and northeastern limbs (i. e., the nonthermal limbs). 
SN 1006 is an ideal target, because its simple bilateral morphology allows us to easily identify regions where particle acceleration is at work and to compare them with regions where electron acceleration does not work efficiently (i. e. the thermal limbs, characterized by a low surface brightness in the radio, X-ray, and gamma-ray bands).
M13 studied the X-ray data by adopting models that assume different electron spectra and found that the loss-limited model developed by \citet{za07} provides the best fit to all the spectra extracted from the nonthermal limbs. Their results show that the loss-limited mechanism is at work in SN 1006 and that radiative losses play a fundamental role in shaping the electron spectrum.

According to the loss limited model, the cutoff frequency in the X-ray synchrotron spectrum, $\nu_0$, depends on the shock velocity and does not depends on the magnetic field, at odds with the time-limited and escape-limited scenarios that predict a dependence of $\nu_{0}$ on $B$. In SN 1006 the shock velocity is expected to be fairly uniform (\citealt{kpl09}, hereafter K9) except in the northwestern limb where the shock is slowed down by the interaction with dense ambient material (\citealt{wgl03}, but see also \citealt{klp13}). Therefore, it is difficult to explain the pronounced small-scale and large-scale variations of $\nu_0$ observed across the SN 1006 shell (e. g., \citealt{mbi09}, \citealt{kpm10}) within the loss-limited framework. We here extend the results presented in M13, by adding two more regions in their spatially resolved analysis of the X-ray spectra. Our final aim is to discuss the possible physical origins of the variations that the cutoff frequency shows across the shell. 

The paper is organized as follows: in Sect. \ref{analysis} we present the results of the X-ray data analysis; in Sect. \ref{disc} we discuss the dependence of the cutoff frequency on the shock velocity and on the electron gyrofactor; and, finally, we summarize our conclusions in Sect. \ref{Conclusions}.

\section{Spectral analysis}
\label{analysis}
We analyze the data of the \emph{XMM-Newton} Large Program of observations of SN~1006 (together with older \emph{XMM-Newton} observations) presented in M13. The data reduction was performed with the Science Analysis System (SAS V12) by adopting the procedures described in detail in \citet{mbd12}. Spectral analysis was performed in the $0.5-7.5$ keV energy band by using XSPEC V12.

To describe the spectra, we here adopt the same absorbed loss-limited model as M13, based on the work by \citet{za07}. The unabsorbed model spectrum has the form 
\begin{equation}
S_X^{ll}\propto \nu^{-2}[1+a(\nu/\nu_0)^{0.5}]^{11/4}\rm{exp}~(-(\nu/\nu_{0})^{0.5})
\end{equation}
where $a=0.38$ and the cutoff energy $h\nu_0$ is a free parameter in the fit. We also include narrow Gaussians to account for residual O VII and O VIII line emission (see M13 for details).

\begin{figure}
\includegraphics[width=\columnwidth]{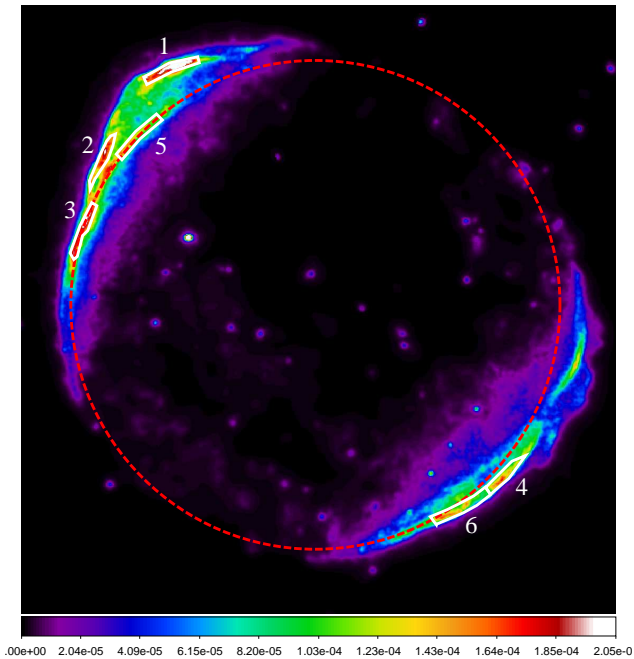}
\caption{EPIC count-rate images of SN~1006 in the $2-4.5$ keV band (bin size $4''$). The regions selected for the spectral analysis of the rim are superimposed. Regions $1-4$ are those analyzed by M13, while we here analyze the spectra extracted from regions 5 and 6. North is up and east to the left. The red circle indicates approximately the shape of the shell and is centered at $\alpha_{J2000}=15^h ~02^m 54.2^2$, $\delta_{J2000}=-41^\circ ~56'19.6''$.}
\label{fig:reg}
\end{figure}

Figure \ref{fig:reg} shows the \emph{XMM-Newton} EPIC image of SN 1006 in the $2-4.5$ keV band. Regions $1-4$ are those analyzed by M13, while we here analyze the spectra extracted from regions 5 and 6. Region 5 is at the same distance from the center of the shell as region 3, but it is closer to the azimuthal center of the northeastern limb. Region 6 is approximately at the same distance from the center as region 4 and is slightly further away from the center than regions 3 and 5. 

In these regions, the loss-limited model provides an accurate description of the spectra and we obtained $\chi^2=3173.7$ (with 3033 d. o. f.) in region 5 and $\chi^2=3079.9$ (with 2874 d. o. f.) in region 6. The best-fit values of the cutoff energy are $h\nu_0=0.286_{-0.006}^{+0.007}$ keV and $h\nu_0=0.310\pm0.009$ keV in regions 5 and 6, respectively.

\begin{figure}
\includegraphics[angle=90,width=\columnwidth]{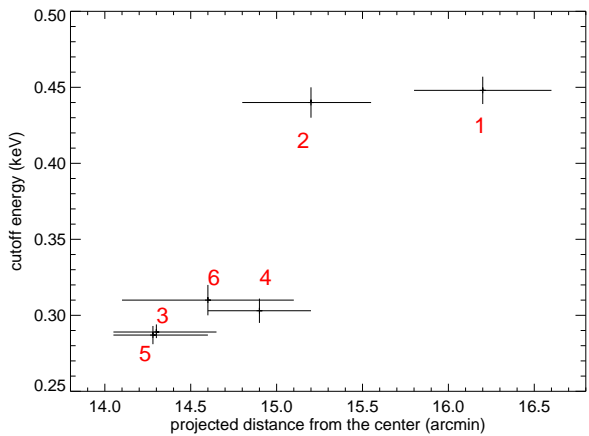}
\caption{Values of the best-fit cutoff energy obtained in regions $1-6$ of Fig. \ref{fig:reg} (the values for regions $1-4$ are from M13) as a function of the distance of the extracting region from the center of the shell (identified as the center of the circle shown in Fig. \ref{fig:reg}). Error bars in the cutoff energy correspond to the $90\%$ confidence level, while the error bars in the x-axis indicate the minimum and maximum projected distance from the center for each region.}
\label{fig:discut}
\end{figure}

Figure \ref{fig:discut} shows how the values of the cutoff energies in spectral regions $1-6$ change with the projected distance of the extracting region from the center of the shell. Regions 3 and 5, located at the same projected distance from the center of the shell, show the same cutoff energy, though they are at different azimuthal angles. The highest values of $h\nu_0$ are reached in regions 1 and 2, that are those at the highest distance from the center of the shell. Both these regions are located in the northeastern limb, where a shock breakout is clearly visible. This breakout can be the result of the propagation of the main shock in a locally rarefied medium.

\section{Discussion}
\label{disc}

In the loss-limited scenario the cutoff energy $E_0$ of the electron spectrum depends on $1/\sqrt{B}$ (e. g., \citealt{rey08}). Therefore the cutoff energy in the corresponding synchrotron radiation (that is proportional to $E^{2}_{0}B$) does not depend on the magnetic field. In particular, it can be shown (\citealt{za07}) that
\begin{equation}
h\nu_0 = \frac{2.2 {\rm keV}}{\eta(1+\sqrt{\kappa})^2} v_{3000}^2 \frac{16}{\gamma_{s}^{2}}
\label{eq:hnu}
\end{equation}

where $\eta$ is the gyrofactor (i. e., the ratio of the electron mean free path to the gyroradius), $\kappa$ is the magnetic field compression ratio, $v_{3000}$ is the shock speed in units of $3000$ km/s, and $\gamma_s$ is the power-law index of particles accelerated at the absence of energy losses. Equation \ref{eq:hnu} shows that $h\nu_0$ increases with the shock velocity. This is in agreement with the general trend shown in in Fig. \ref{fig:discut}, which seems to suggest that (in the nonthermal limbs) $h\nu_0$ increases with the distance from the center. 

K9 have shown that the proper motion of the SN 1006 forward shocks is almost invariant over the whole northeastern limb ($v\sim5000$ km$/$s). They also found that there is no significant velocity difference between the inner shock (our region 5) and the outer shock (regions 1 and 2 of Fig. \ref{fig:reg}). Nevertheless, the significantly higher distance from the center clearly indicates that the outer shock must have experienced a higher shock velocity than the inner shock during the remnant evolution. We can therefore interpret the high values of $h\nu_0$ measured in regions 1 and 2 as the result of a local enhancement in the shock speed.

On the other hand, the azimuthal profile of the cutoff frequency across the rim of SN 1006 shows variations of more than one order of magnitude. These variations have been observed by adopting the SRCUT model, but similar results, in terms of the value of $h\nu_0$, can be obtained with the loss-limited model (see M13). In particular, in the nonthermal limbs the values of $h\nu_0$ are systematically higher than in the thermal limbs (see \citealt{rbd04} and \citealt{mbi09}). This result can hardly be explained as an effect of fluctuations in the shock velocity only, since unrealistic fluctuations by a factor $>3$ should be invoked.

Another possibility is that the variations of $h\nu_0$ are associated with different values of the gyrofactor. The loss-limited model adopted here assumes Bohm diffusion\footnote{In general, the shape of the cutoff in the X-ray spectrum  depends on the nature of the diffusion process, as shown by \citet{bla10}}, that corresponds to $\eta=1$. 
As first suggested by \citet{za07}, one can use Eq. \ref{eq:hnu} to obtain estimates of deviation of the diffusion process from the extreme Bohm condition, when the velocity of the SNR shock front is known. If we assume $v=5000$ km$/$s (as measured by \citealt{kpl09}) and we put $\gamma_s=4$ and $\kappa=\sqrt{11}$ (that is, the increase in the isotropic random $B$ field for a shock compression ratio $r=4$), we obtain $\eta = 1.7$ for $h\nu_0=0.448$ keV (i. e. the value obtained by M13 in region 1, see also Fig. \ref{fig:discut}). This result clearly indicates that in region 1 the electron diffusion proceeds in a regime that is very close to the Bohm limit. This conclusion does not change if we assume that in the recent past the shock velocity in region 1 has been higher than 5000 km/s, given that $\eta$ depends only on the square root of $v$. For example, we would have obtained $\eta=2.5$, by assuming $v=6000$ km$/$s.
In regions 3 and 5 (where $h\nu_0 = 0.29$ keV, see Fig. \ref{fig:discut}), we still obtain a low value for the gyrofactor $\eta = 3.8$. There are no measures of the shock proper motion in the southwestern limb of SN 1006, but by adopting the realistic guess $v=5000$ km$/$s, we obtain $\eta\sim3.6$ therein.
Even taking into account possible variations in the shock speed, we can then conclude that in both nonthermal limbs the electron diffusion is close to the Bohm conditions.

It is possible that the low values of the cutoff energy measured in the southeastern and northwestern thermal limbs (\citealt{mbd12,mbi09}) are associated with a much higher value of the gyrofactor and of the electron mean free path in these regions.
However, we point out that the spectral analysis carried in Sect. \ref{analysis} and in M13 has proven that the loss-limited model is at work in the nonthermal limbs. It is not possible to accurately study the shape of the X-ray synchrotron emission in thermal limbs because of the dominant contribution of thermal emission from shocked ejecta and shocked ambient medium (the latter being significant at high energies, as shown by \citealt{mbd12}). Therefore, we cannot exclude that a mechanism different from the radiative losses limits the maximum electron energy in the thermal limbs of SN 1006.
In any case, a dependence of particle injection and$/$or acceleration (and diffusion) on the shock obliquity must be invoked to explain the observed low values of the cutoff energy in the thermal limbs.

\section{Conclusions}
\label{Conclusions}
Different studies have shown that the roll-off frequency of the X-ray synchrotron emission is significantly higher in the northeastern rim of SN 1006 than in its southwestern rim (e. g. \citealt{bfh08}). As shown in Fig. \ref{fig:discut}, this difference originates in the outer northwestern shock only (regions 1 and 2 in Fig. \ref{fig:reg}), where $h\nu_0$ reaches its highest values.
This result is in agreement with expectations, since, as shown by M13, the loss-limited scenario is at work in the nonthermal limbs of SN 1006, and it predicts that the cutoff energy increases with the square of the shock speed (and the outer northeastern shock is the region located at the highest distance from the center of the shell). We therefore conclude that the hardest X-ray emission of the northeastern limb can be an effect of the high shock speed in its outer shock.

We also verified that in both nonthermal limbs the electron diffusion proceeds at a regime that is very close to the Bohm limit, with values of the gyrofactor in the range $\sim 1.5-4$.
The low values of $h\nu_0$ in thermal limbs can instead be due to a much larger deviation from the Bohm conditions ($\eta>>1$). However, it is not possible to exclude that in thermal limbs both the acceleration process and the mechanism that limit the energy achieved by the electrons are different from those that are at work in nonthermal limbs, possibly because of the different shock obliquity.



\acknowledgements
This paper was partially funded by the ASI-INAF contract I$/$009$/$10$/$0. M. M. thanks Rino Bandiera for interesting discussions and suggestions.

\newpage



\end{document}